\documentclass[aip, msmath,amssymb,reprint]{revtex4-2}

\usepackage[T1]{fontenc}
\usepackage[utf8]{inputenc}

\usepackage{booktabs}
\usepackage{threeparttable}
\usepackage[d]{esvect}
\usepackage{verbatim}
\usepackage[version=4]{mhchem} 
\usepackage{float}
\usepackage[normalem]{ulem}

\usepackage{graphicx}
\usepackage{hyperref}
\hypersetup{
  colorlinks,
  citecolor=blue,
  linkcolor=blue,
  urlcolor=blue
}
  
\usepackage{siunitx}
\DeclareSIUnit\bar{bar}
\DeclareSIUnit\angstrom{Å}
\DeclareSIUnit\bohr{\text {b}}
\DeclareSIUnit\calorie{cal}

\usepackage{textgreek}
\usepackage{chemmacros}
\usepackage{amssymb}

\usepackage{float}

\begin{document}

\title{Bottlenecks in Hamiltonian-Adaptive Resolution Simulation Method for Modeling Interfaces}

\author{Hari Haran Sudhakar}
\affiliation{Sorbonne Universit\'e, CNRS, Physicochimie des \'Electrolytes et Nanosyst\`emes Interfaciaux, F-75005 Paris, France}

\author{Alessandra Serva}
\email{alessandra.serva@sorbonne-universite.fr}
\affiliation{Sorbonne Universit\'e, CNRS, Physicochimie des \'Electrolytes et Nanosyst\`emes Interfaciaux, F-75005 Paris, France}
\affiliation{R\'eseau sur le Stockage Electrochimique de l'Energie (RS2E), FR CNRS 3459, 80039 Amiens Cedex, France}

\author{Rocio Semino}
\email{rocio.semino@sorbonne-universite.fr}
\affiliation{Sorbonne Universit\'e, CNRS, Physicochimie des \'Electrolytes et Nanosyst\`emes Interfaciaux, F-75005 Paris, France}

\date{\today}

\begin{abstract}
The Hamiltonian-Adaptive Resolution Simulation (H-AdResS) method allows to combine atomistic and particle-based coarse-grained models in a single simulation box, which makes it very attractive to model systems containing interfaces or reactive regions surrounded by an interacting environment. In our previous work [J. Chem. Inf. Model. (2026) 66 (15): 9215–9225], we implemented H-AdResS in LAMMPS (2 Aug 2023 version) and extended its use to interfaces, focusing on MOF/CO$_2$ interfaces as an example. We found that, despite its advantages, using this method properly for this kind of systems is not trivial. In this work, an in-depth analysis of the impact of the choice of thermostatting schemes and long-range electrostatics models is presented. Even though its Hamiltonian formulation enables performing H-AdResS simulations within constant temperatures ensembles, not every thermostat is appropriate. We demonstrate that Langevin thermostat is a reliable choice for this method, while Nosé-Hoover results in artifacts. In addition, we show that using short-range models such as the Damped Shifted Force method for electrostatics, a popular choice for H-AdResS simulations, can lead to non-physical results when modeling interfaces. The need of capping strategies to deal with discontinuities in forces and energies arising from abrupt changes in resolution is also discussed. Finally, the impossibility of changing the definition of the H-AdResS Hamiltonian to include a gradual interpolation of the bonded degrees of freedom is discussed. We hope that this contribution helps the reader to appropriately set up H-AdResS simulations and to assess if this method can be used to accurately model their system of interest.  
\\


\end{abstract}

\maketitle

\section{Introduction}

Atomistic simulations make it possible to study the structure and dynamics of phenomena happening at 1-100 nanoseconds timescales in systems of characteristic lengthscales of a few nanometers, which often covers molecular scales. However, modeling low concentration or inhomogeneous systems including interfaces, may require larger samples. Adaptive resolution simulations (AdResS \cite{Praprotnik2005}) were created with the objective of being able to model large systems while keeping atomistic (AA) resolution in a region of the simulation box. The rest of the system is typically modeled in a less computationally costly, lower resolution particle-based coarse grained (CG) model. Within this method, molecules are free to diffuse from one region to another, and their resolution is seamlessly adapted\cite{Praprotnik2008}.

Initially, a hybrid region, in which molecules possess a dual AA/CG identity, was placed between the atomistic and coarse grained box regions, although more recent versions of the method without this region were also developed \cite{Gholami2021, Thaler2020, Krekeler2018}. Non-bonded forces over the molecules in the hybrid region are interpolated from the atomistic and CG force fields, while bonded forces are always computed according to the atomistic force field, even in the region of the simulation box that is reserved to the CG model \cite{Wang2012}. Conversely, in the Hamiltonian-Adaptive Resolution Simulation (H-AdResS) method, energies are interpolated in the hybrid region\cite{Potestio2013, Espaol2015}. This makes the method compatible with constant temperature ensembles. 

Within the past $\sim$15 years since their creation, AdResS and H-AdResS methods have been used to study a wide variety of bulk liquid systems \cite{Praprotnik2006, Wang2018, Krekeler2017, DuenasHerrera2026, ShadrackJabes2018}. Recently, we have tested the applicability of H-AdResS to model interfaces between a condensed and a gas phase \cite{Sudhakar2026}. Up to that moment, the H-AdResS method was implemented in a single LAMMPS version \cite{Heidari2016} from the year 2016. Since then, LAMMPS has had significant modifications, such as an update to the most recent version of C++ 11, new functions and utilities, and the addition of additional fix/compute styles. Because of this, it is not feasible to directly transpose the 2016 H-AdResS implementation to the more recent LAMMPS versions. In our previous work, we provided a new implementation of H-AdResS in LAMMPS 2023.\cite{Sudhakar2026} We benchmarked our implementation against the implementation from 2016 and a stable LAMMPS release and we found that they were in very good agreement (see Figure \ref{fig:rdf-water}). Therefore, all the H-AdResS simulations reported in this work are performed using our implementation.

When setting up our H-AdResS simulations, we faced a number of difficulties. These include numerical issues when molecules entered the hybrid region from the CG one and when the sampling of the canonical ensemble was done through the Nosé-Hoover \cite{Nos1984} equations. Moreover, when extending this method to modeling charged interfaces, the way in which charges are treated in the force field needed to be revised. In this article, we discuss these aspects of H-AdResS simulations, in order to guide the readers that are starting in the field to correctly set up their H-AdResS simulations.

Previous AdResS and H-AdResS works mention that the inclusion of the CG bonded degrees of freedom in the CG region is a desirable extension to the method.\cite{Junghans2010, Praprotnik2011, esp++} Note that it is possible to parameterize CG force fields considering the underlying AA degrees of freedom,\cite{KarimiVarzaneh2008, diPasquale2012, DiPasquale2014,  Gowers2015} but this is not a common practice, may require additional parametrization and has been only proven for a small set of systems. We have attempted including CG bonded terms in the Hamiltonian within our LAMMPS implementation, but found issues that could not be solved and that we discuss in detail. Briefly, since H-AdResS implementations in LAMMPS work with a fixed number of particles (the maximum number, that is the atomistic number of particles), without bonded terms to maintain the underlying atomistic topology, numerical errors that cannot be fixed arise. If the number of particles was allowed to change on-the-fly, numerical issues may still appear due to the introduction of atomistic degrees of freedom in an orientation that is not necessarily compatible with other species in the hybrid region.

This article is organized as follows. First, the methods are summarized, together with the characteristics of the thermostatting schemes tested, the electrostatics treatments and simulation details. Section III contains our results and discussion, while the final section summarizes our conclusions.

\section{Methods}

In concurrent coupling frameworks, particles actively diffuse between regions differing in resolution through a hybrid region where the interactions are interpolated.\cite{Praprotnik2005} For H-AdResS, the global Hamiltonian is given by:\cite{Potestio2013}
\begin{equation} 
H  = \kappa + U_{bonded} + \sum_a \{\lambda_a U^{AA}_a + (1-\lambda_a) U^{CG}_a \}
\label{eq:hamiltonian}
\end{equation} 
where $\kappa$ is the total kinetic energy, $U_a^{AA}$ and $U_a^{CG}$ are the non-bonded potential energies of molecule $a$ in atomistic (AA) and coarse-grained (CG) resolution respectively and $U_{bonded}$ contains the potential energy of every bond, angle, and proper and improper dihedral atomistic degrees of freedom. They are integrated throughout the simulation box, including for the molecules in the CG region at every simulation step. $\lambda_a$ denotes the resolution of molecule $a$, and can take values between 0 (CG) and 1 (AA). In the hybrid region, $\lambda_a$ changes smoothly from 0 to 1 or vice versa. Here, both AA and CG non-bonded interactions are calculated and scaled according to $\lambda_a$ values. 
The force acting on atom $i$ of molecule $a$ is given by,
\small
\begin{equation}\label{eq:force}
\begin{split}
F_{ai} &= F_{ai}^{bonded} + \sum_{\substack{b,\,b \neq a}} 
\left\{ \frac{\lambda_a + \lambda_b}{2} F_{ai|b}^{AA} + \left(1 - \frac{\lambda_a+ \lambda_b}{2}\right) 
F_{ai|b}^{CG} \right\} \\
&\quad - \left[U_{a}^{AA} - U_{a}^{CG}\right] \nabla_{ai} \lambda_a.
\end{split}
\end{equation}
\normalsize

The term $ F_{ai}^{bonded}$ represents the force due to the bonded interactions and the second term represents the force experienced by atom $i$ of bead $a$ due to its non-bonded interactions with bead $b$. $F_{ai}^{drift} = - \left[U_{a}^{AA} - U_{a}^{CG}\right] \nabla_{ai} \lambda_a $ introduces a drift force in the hybrid region. Since AA and CG regions are governed by different force fields, this can result in inhomogeneous pressure and densities along the simulation box. To ensure uniform pressure and density, the Hamiltonian is modified by introducing compensation terms:\cite{Heidari2016}
\begin{equation}
H' = H - \sum_{a=1}^{N} \Delta H(\lambda(X_{a})).
\label{eq:modified-hamiltonian}
\end{equation}

Two different compensation routines are used to ensure uniform pressure and density respectively. Uniform pressure is attained by removing the average drift force in the hybrid region as discussed above. For uniform density, a thermodynamic force correction is applied on the molecules in the hybrid region,\cite{Fritsch2012} as shown in eqn (\ref{eq:modified-hamiltonian}).

\subsection{Thermostats}
In MD simulations, temperature is given by the expression\cite{temp}:
\begin{equation}
    T = \frac{2E_{kin}}{N_{dof}k_b}
    \label{eq:Temp-eqn}
\end{equation}
where $N_{dof}$ denotes the number of degrees freedom, $k_b$ the Boltzmann constant and $E_{kin}$ is the instantaneous kinetic energy:
\begin{equation}
    E_{kin} = \frac{1}{2}\sum_{i=1}^{N} m_iv_i^2
\end{equation}
with $m_i$ and $v_i$ the mass and velocity of particle $i$ respectively. The average target kinetic energy of an unconstrained system is given by:
\begin{equation}
    \langle E_{kin}\rangle = \frac{3}{2}Nk_bT
\end{equation}
where N denotes the total number of particles. Newton's equations of motion are integrated to evolve the system in time which allows us to sample the NVE ensemble. However, most experiments are carried out in constant temperature conditions, which corresponds to a constant temperature ensemble, such as the canonical ensemble. In order to maintain the temperature constant, various thermostatting algorithms were developed. Thermostats used in MD simulations can be broadly classified into two classes: global and local thermostats.\cite{Shiraishi2026} 

One of the most commonly used global thermostat is the Nosé-Hoover (NH) thermostat\cite{Hoover1985,Nos19841}. It is a deterministic thermostatting algorithm which couples the whole system to an imaginary heat bath, denoted by the variable $\xi$. Thus, the modified equations of motion become $\dot{p_i} = F_i -\xi p_i$, where $p_i$ denotes the momentum of particle $i$ and $\xi$ is directly proportional to the difference between instantaneous and target kinetic energy, given by the expression:
\begin{equation}
    \frac{d\xi}{dt} = \frac{1}{M}\left \{ \frac{1}{2}\sum_{i=1}^{N} m_iv_i^2 - \frac{3N+1}{2}k_bT \right \}
\end{equation}

The factor is 3N+1 instead of 3N, because of the addition of the new degree of freedom $\xi$, and $M$ denotes the damping parameter of the dynamics of the heat bath $\xi$. Since all the particles are scaled uniformly, the effect of the thermostat on inter-particle forces is low. This makes it a good choice for minimizing perturbations over the system.

On the other hand, there are local thermostats such as the Langevin (LN) thermostat\cite{Brnger1984}, which couples each particle to a virtual heat bath independently. It is a stochastic thermostatting algorithm, as it introduces random forces and local friction to individual particles. The modified equation of motion becomes $\dot{p_i} = F_i - \gamma m v_i + R_i $. $R_i$ is the random force term and $\gamma$ is the friction. Since every particle is independently coupled to a virtual heat bath, local thermostats guarantee thermalization even in the absence of particle collisions (implicit solvent cases).\cite{Hicks2021} A discussion on the choice of the right thermostat for H-AdResS simulations is provided in the Results section.

\subsection{Electrostatics}

Electrostatic interactions decay at the rate of $1/r$ which is much slower when compared to dispersion interactions that decay at $1/r^6$. Since the decay of $1/r$ is much slower, electrostatic interactions are essentially long-ranged. In periodic systems, conventional MD simulations handle long-range electrostatic interactions using Ewald summation methods such as Particle-Particle Particle-Mesh (PPPM) and Particle-Mesh Ewald (PME) \cite{Hockney2021}. These methods accurately capture the long-range nature of Coulombic interactions by dividing them into short- and long-range components. The first one is calculated in the real space as the Coulombic interaction between the charge pairs within a defined cutoff distance. The long-range component is calculated in the reciprocal space, as an interaction of a particle with all the other particles present in the box and their periodic images, using Fourier Transformations (FTs).

Though these approaches are robust, they are not compatible with adaptive resolution methods.\cite{Heidari2018, heidari2019} Long-range components are calculated using FTs by considering the position of all the charges in the simulation box. However, the active degrees of freedom and the physical properties vary across the simulation box in H-AdResS simulations due to changes in resolution. This makes global techniques like FTs inconsistent. To circumvent this, early adaptive resolution implementations (AdResS) used the Reaction Field (RF) method \cite{Brooks1989,Barker1973}. The latter treats the local charges explicitly for the short-range part, while all the molecules beyond the cutoff $R_C$ are treated as a homogeneous dielectric medium with a constant permittivity ($\epsilon$).

Another method that emerged as a powerful alternative to RF is the Damped Shifted Force (DSF) method \cite{Fennell2006}. Coulombic interactions calculated by DSF method are given by\cite{DSFpotential}:

\small
\begin{equation}
\begin{aligned}
E_{ij}= q_i q_j \Bigg[
&\frac{\operatorname{erfc}(\alpha r)}{r}
-\frac{\operatorname{erfc}(\alpha r_c)}{r_c} \\
&+
\left(
\frac{\operatorname{erfc}(\alpha r_c)}{r_c^2}
+\frac{2\alpha}{\sqrt{\pi}}
\frac{\exp(-\alpha^2 r_c^2)}{r_c}
\right)
(r-r_c)
\Bigg]
\end{aligned}
\end{equation}
\normalsize

where $r_c$ is the cutoff distance for the pair wise calculations between charges $qi$ and $q_j$. DSF computes electrostatic interactions purely from pairwise calculations, effectively reducing them to a finite range potential, without assuming a dielectric medium. A damping parameter ($\alpha$) is introduced to smoothly nullify the Coulomb potential as the distance between the charges increases. This damped potential gives rise to a screening effect. The parameters $\alpha$ and $r_c$ can be tuned to reproduce benchmark structural properties.

Though both RF and DSF are finite range methods and computationally efficient, RF works on the assumption of a uniform medium beyond the cutoff and this assumption does not stand in the case of interfaces or complex biomolecular systems, where local dielectric properties fluctuate significantly.\cite{Fukuda2012} If the dielectric constants are not accurate enough, RF can result in incorrect energies and forces. \cite{vanderSpoel1998} In systems like salt solutions, RF requires knowing the Debye length ($\kappa$), to model the ionic screening from the beginning of the simulation. \cite{Allen2017, Bevc2013, Barker1994} On the other hand, DSF reproduces the screening effects as an emergent property and does not require a dielectric constant. Moreover, its pairwise form makes it fit into the H-AdResS Hamiltonian seamlessly. This enables the effective handling of varying resolution and ionic environments.\cite{heidar2018}

Electrostatic interactions play a crucial role in determining the structure of double layers and local potential gradients at interfaces, which directly governs ion transport and energy barriers for charge transport. Thus, accurate estimations of these interactions are necessary for the robust modeling of electrochemical devices such as batteries and supercapacitors. \cite{Becker2023,Scalfi2021}

\section{Simulation details}

Liquid water, ZIF-8 and ZIF-8 loaded with \ce{CO2} gas were simulated within the H-AdResS scheme with Nosé-Hoover and Langevin thermostats to sample the NVT ensemble at 300 K, using a modified LAMMPS 2023 (2 Aug 2023 - Update 2) \cite{Thompson2022} code that includes the possibility of running H-AdResS presented in our previous work.\cite{Sudhakar2026} All the simulations were performed with a timestep of 1 fs. For all the reference AA simulations, long-range interactions were calculated using the particle-particle particle-mesh (PPPM) method\cite{pppm}, with a relative error of $10^{-6}$.  Nosé-Hoover thermostat with damping constant of 0.1 ps and barostat with damping constant of 1 ps were used. All H-AdResS simulations were started without compensations. Both density and pressure compensation forces were updated every 1 ps and a damping constant $\alpha$ of 0.2 was used for the DSF potential.

10240 water molecules were filled in a rectangular box of dimensions $200$ \AA \ x $40$ \AA \ x $40$ \AA \ using the \textit{create\textunderscore atoms} function in LAMMPS. The SPC/E water model\cite{Berendsen1987} was used for AA resolution and the bonds and angles are constrained using SHAKE algorithm \cite{shake}, while the Weeks-Chandler-Andersen (WCA) potential\cite{Weeks1971} was employed for the CG model. The initial configuration generated by LAMMPS was equilibrated in the NPT ensemble at 300 K and 1 atm for 10 ns. Then, a 10 ns production run was performed in the NVT ensemble at 300 K. For H-AdResS simulations, an equilibrated configuration generated from an AA simulation was used as initial configuration. The size of the rectangular AA and hybrid regions were set to be 60 \AA\ and 25 \AA \ in the $x$ direction (the direction perpendicular to the change in resolution), respectively. After a 50 ps-long run, either one or both compensations were activated. The equilibration was performed for 500 ps. Then, a 10 ns production run was performed. For pressure compensation, $\nabla\lambda$ was set to be 0.02 and for density compensation, bin width was set to 1.5 \AA. A Langevin thermostat with 0.1 ps damping constant was used for the whole box.

To measure the dipolar orientation profile, a slab of liquid water placed in the center of a rectangular box in contact with vacuum in the x-direction was simulated. We measured the average dipolar orientation $\langle cos(\theta) \rangle$, a crucial property governed by electrostatics, where $cos\theta_i$ denotes the cosine of the angle between the dipole vector of the water molecule $i$ and the vector normal to the interface ($x$ axis). Three different thickness of water slab were simulated: 25 \AA, 30 \AA \ and 35 \AA. All three slabs were simulated using either the PPPM or the DSF method, as well as using three different LAMMPS implementations: Stable release (22 Jul 2025 - Update 1) \cite{Thompson2022}, H-AdResS 2023 (2 Aug 2023 - Update 2) \cite{Sudhakar2026} and H-AdResS 2016 (28 Apr 2016 - ICMS) \cite{Heidari2016}. The effect of the vacuum size over the 25 \AA \ water slab was also investigated using three different vacuum sizes namely 40 \AA \, 60 \AA \ and 80 \AA \ respectively. The SPC/E water model\cite{Berendsen1987} was used and the bonds and angles are constrained using SHAKE algorithm \cite{shake}. For the all configurations, the equilibration was performed for 10 ns in the NPT ensemble. Then, a 10 ns production run was performed in the NVT ensemble. This extensive analysis was performed to understand the effect of different long-range methods, slab sizes, vacuum sizes and LAMMPS implementations in capturing the dipolar orientation profile of the water interface.

\begin{figure}
    \centering
    \includegraphics[width=0.5\textwidth]{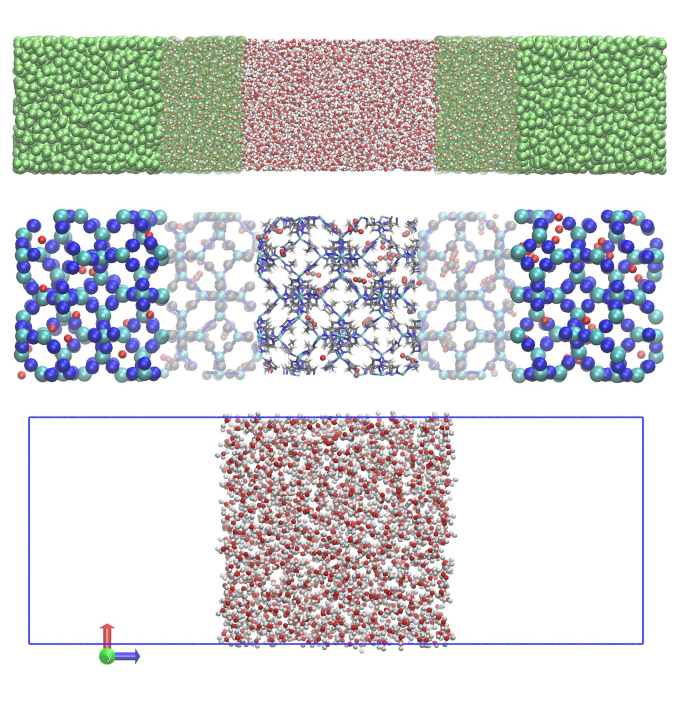}
    \caption{Schematic representation of (from top to bottom) bulk liquid water, ZIF-8 loaded with \ce{CO2} in a H-AdResS scheme, and water/vapour interface used to study the dipolar orientation profile from simulations performed by treating electrostatics via either PPPM or DSF methods.}
    \label{fig:snapshot}
\end{figure}

For the ZIF-8/\ce{CO2} system, the interactions in the AA region were governed by ZIF-FF \cite{Park2006} and \ce{CO2} was modeled using the EPM-2 model.\cite{ZHU2009} For both ZIF-8 and \ce{CO2}, a MARTINI based force field was used in the CG region. \cite{Alvares2025} In the CG scheme, ZIF-8 is made up of two beads \textit{i.e.} the whole ligand as one bead and Zn as another bead, while \ce{CO2} molecules are represented by one CG bead each, as shown in Figure S1. A 8x2x2 supercell of size $ 135.68$ \AA \ x $33.92$ \AA \ x $33.92$ \AA \ was studied with 40 \AA \ of AA and 25 \AA \ of hybrid regions in the x-direction respectively. \ce{CO2} gas was loaded into the MOF using the Grand Canonical Monte Carlo (GCMC) method (\textit{fix gcmc} command in LAMMPS). AA and CG force fields reproduce an identical loading of 1.5-2 \ce{CO2} molecules per unit cell \cite{Sudhakar2026}, which is in agreement with the experimental observations (0.8 to 1.8 molecules of \ce{CO2} per unit cell of ZIF-8). \cite{NeumannBarrosFerreira2024, Klomkliang2025, Wu2014, Zhang2014, Pusch2012} The initial configuration was generated using a python script and equilibrated in the NPT ensemble at 300 K and 1 atm. Then, a 5 ns production run was performed in the NVT ensemble at 300 K. For H-AdResS simulations, the size of AA and hybrid regions were set to be 40 \AA \ and 25 \AA \ respectively. After a 50 ps-long run, compensations were activated. The equilibration was performed for 500 ps. Then, a 5 ns-long production run was performed. For pressure compensation, $\nabla\lambda$ was set to 0.05 and for density compensation, bin width was set to 2.0 \AA. Two Langevin thermostats, one for AA+hybrid regions and another one for the CG region, with 1 ps damping constant, were used. 

\section{Results}

In the following sections, we first discuss how to set up thermostats in H-AdResS simulations correctly. We analyze the Nosé-Hoover and Langevin thermostatting strategies and whether one or more thermostats are needed. Secondly, we analyse the suitability of the DSF method for electrostatics to simulate interfaces within the H-AdResS scheme. Next, we detail force and energy issues that can arise at the interface between the coarse-graining and the hybrid region, and propose a strategy to solve them. Finally, we discuss the possibility of including the CG bonded contributions in the H-AdResS Hamiltonian. 

\subsection{Nosé-Hoover or Langevin thermostat?}

\begin{figure}[b]
    \centering
    \includegraphics{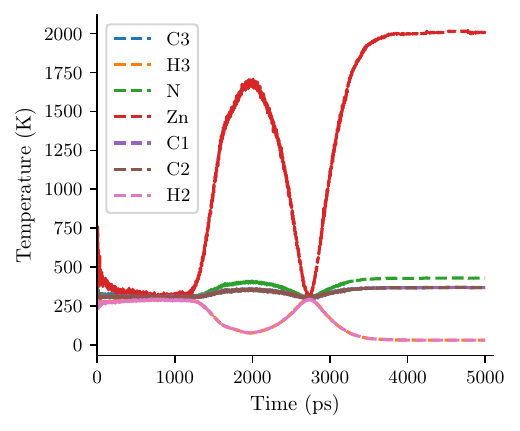}
    \caption{Temperature associated to different atom-types of ZIF-8 (see supporting information) in the AA region within the H-AdResS scheme with both compensations turned on and using the Nosé-Hoover thermostat.}
    \label{fig:water_temp}
\end{figure}

Bulk ZIF-8 was simulated (i) using three different LAMMPS implementations: Stable release (22 Jul 2025 - Update 1) \cite{Thompson2022}, our implementation H-AdResS 2023 (2 Aug 2023 - Update 2) \cite{Sudhakar2026} and H-AdResS 2016 (28 Apr 2016 - ICMS) \cite{Heidari2016} with the NH thermostat and (ii) using our implementation (H-AdResS 2023 (2 Aug 2023 - Update 2) \cite{Sudhakar2026}) with NH thermostats varying the damping constant. In both cases, the temperatures of different atom-types of ZIF-8 (see figure \ref{fig:mapping}) were measured and reported in the SI (see figures \ref{fig:damping} and \ref{fig:version-thermo}). Across all three LAMMPS versions and damping constants tested, the per-atom type temperatures remain consistent with each other and with the target temperature. This indicates that neither the specific LAMMPS implementation nor the choice of NH damping constant introduces any systematic bias in the thermal equilibration of ZIF-8.

For H-AdResS simulations, the temperatures associated with different ZIF-8 atom-types in the AA region were measured from simulations using the NH thermostat and are presented in Figure \ref{fig:water_temp}. An uneven distribution of temperature between different species present in the simulation box is observed (see figure \ref{fig:aa-vs-hars}), though the global temperature is at the set temperature of 300 K. Global thermostats work on the assumption of the equipartition theorem that collisions between particles will naturally transfer energy\cite{Arabzadeh2026}. The NH thermostat measures the total kinetic energy of the system and scales the velocities of all the particles uniformly, as previously discussed. However, in the H-AdResS scheme, the simulation box is divided into different regions and the interactions within the regions and between the regions are governed by different equations of state. An overall global scaling factor, such as that used in the NH thermostat, cannot compensate the imbalance in the local temperatures. This results in non-physical thermal gradients among the atomic species.

\begin{figure}[t]
    \centering
    \includegraphics{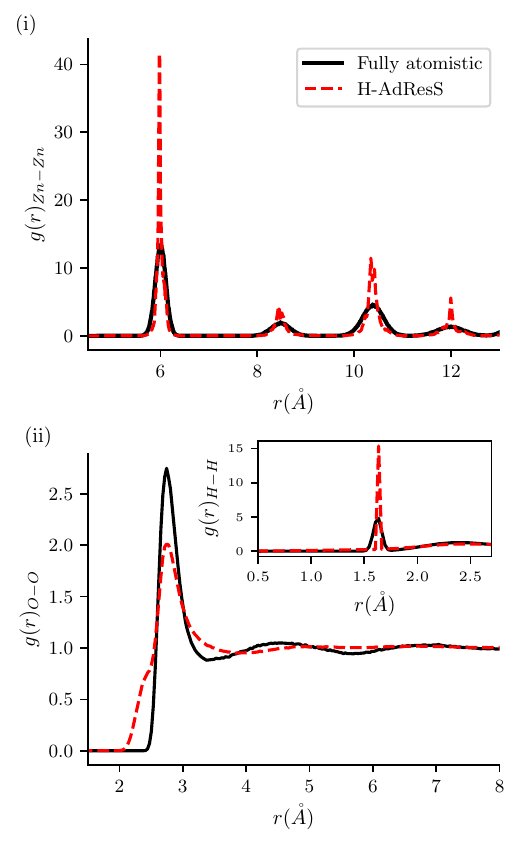}
    \caption{Radial distribution function, g(r), of (i) Zn--Zn pairs for solid ZIF-8 and (ii) O--O pairs and H--H pairs (in the subplot) for liquid water in the AA region of the H-AdResS scheme (red long dashes) compared to the reference fully atomistic simulations (black). Both simulations were performed using the Nosé-Hoover thermostat.}
    \label{fig:rdf-plot}
\end{figure}

The fact that the temperature is not correctly accounted for different atom-types induces a wrong structure in the system. The top panel of Figure \ref{fig:rdf-plot}(i) represents the Zn -- Zn radial distribution function (g(r)) for ZIF-8. Compared to the reference fully atomistic simulation (performed using the NH thermostat), the peaks are very thin and sharp. A similar trend is observed in the bottom panel of Figure \ref{fig:rdf-plot}(ii), for the O--O and H--H g(r)s for liquid water, that are the result of similar non-physical thermal gradients. When visualizing the simulation trajectories of solid ZIF-8, intermittent freezing and flying ice cube effect\cite{Braun2018} were observed.

This problem was solved by using the Langevin thermostat instead of the NH thermostat in previous AdResS and H-AdResS studies. \cite{Wang2018, Potestio2014} Atomistic and coarse-grained regions retain different numbers of active degrees of freedom and different potential energy surfaces. A local Langevin thermostat forces energy equipartitioning on a per-particle basis, by applying random forces and friction to individual particles independently. This ensures that the particles strictly tend towards the target temperature irrespective of their resolution at a given time and guarantees a uniform temperature distribution across the species. 

Even though the LN thermostat solves the artifact in temperature/kinetic energy  (see Figure \ref{fig:two-ln}), it still poses some challenges that may affect the physics of the system being simulated. The term $R_i$ introduces random forces in the form of Gaussian white noise, which are applied on every particle independently. Phenomena such as collective oscillations resulting in phonons traveling through a solid, local correlations between neighboring atoms or the flow of a liquid around a solute could be disrupted by these random forces.\cite{Goga2012} 

When using the LN thermostat, we apply a continuous local friction force. This friction force (drag) can slow down the evolution of the particles in time and space. A large coupling time can be used, since the coupling time is inversely proportional to the friction force. But a choice of very large coupling time may lead to incorrect sampling of events that do not happen on a similar timescale. A strategy to choose the right coupling time is running multiple fully atomistic simulations using the LN thermostat with varying coupling times. Then, a dynamic property such as the velocity auto-correlation function can be measured and compared with that computed from a fully atomistic NVE simulation. By doing this, we can select a coupling time that minimizes the perturbation caused by the friction force. \cite{Carof2013}

In H-AdResS, the main computational advantage comes from the CG region, where dynamics is faster due to the reduced degrees of freedom, and the potential energy surface is smoother. When a LN thermostat applies a uniform friction on all the particles across the simulation box, it effectively introduces drag in the CG region. This can eliminate key benefits of using a CG model, such as faster phase-space exploration and computational efficiency. To prevent this, more than one LN thermostat with different coupling times for different regions can be set up. The coupling time for the different thermostats can be selected following the strategy discussed above.

While simulating a dual resolution system with a single thermostat, the average temperature of the system may appear constrained but the temperature of different groups may diverge\cite{Gowers2015} or cause spurious fluctuations (see Figure \ref{fig:one-ln}). When the systems were simulated with two LN thermostats instead, no non-physical thermal gradients were found. The H-AdResS scheme reproduced the structure and dynamic properties (see Figure \ref{fig:two-ln}), in good agreement with the reference fully atomistic simulations performed using the NH thermostat.\cite{Sudhakar2026}

To sum up, the NH thermostat is not a recommended choice for H-AdResS simulations. Instead, a combination of two LN thermostats with different damping constants for different regions is a sensible choice. Care should be taken while choosing the right damping constant to ensure the correct dynamics in the AA region within the H-AdResS scheme.

\subsection{Electrostatics}

Another crucial aspect in setting up H-AdResS simulations involving interfaces, is the proper treatment of long-range interactions. To tackle this aspect, we studied a water-vapour interface using the DSF method, currently used for H-AdResS simulations. This system was selected as a prototype of a liquid-vapor interface where the simplicity of the structure is combined with the necessity of a proper treatment of electrostatics, to reproduce the orientation of dipoles at the interface. Here, the dipolarorientation profile is expected to tend to zero in the bulk water region with distinct positive and negative peaks at the interface \cite{Becker2023}. 
\begin{figure}
    \centering
    \includegraphics{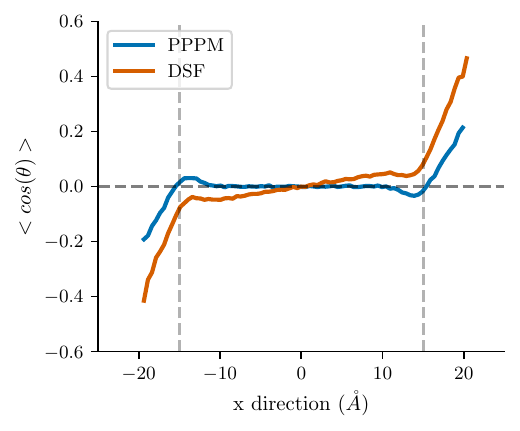}
    \caption{Dipolar orientation function, $\langle cos(\theta) \rangle$, plotted against the direction perpendicular to the liquid-vapour interface (x), obtained from fully atomistic simulations performed using the PPPM (blue curve) and DSF (orange curve) methods to treat electrostatic interactions. The vertical dotted lines denote the position of the liquid water slab. Both blue and orange curves are computed using the same python code. \cite{maicos}}
    \label{fig:short_vs_long}
\end{figure}

In Figure \ref{fig:short_vs_long}, we compare the dipolar orientation profile of the liquid water slab in contact with vapour, computed using either a long-range (PPPM) or the DSF method. The peaks at the extremes of the blue curve indicate the preferential orientation and layering of water molecules at the interface, a well-known behavior for this system. \cite{Klauda2007, Feller1996} Far from the interface, the curve tends to zero, denoting no preferential orientation in the bulk phase. The structure and polarization at the interface are well captured by the PPPM method. Contrarily, the estimation made using DSF method is found to be poor (orange curve) and the interfacial structure deviates significantly. We also performed the analysis on water slabs of (i) different thickness and different vacuum region sizes (see Figure \ref{fig:dipole-profiles}) using our implementation \cite{Sudhakar2026} and (ii) using three different versions of LAMMPS (see Figure \ref{fig:my_label}), respectively. In all cases, the tendency of the curves was found to be same, thus suggesting that the problem is not rooted in the system setup but in the method itself. This suggests that properties for which long-range interactions are crucial, such as electric double layers, global charge asymmetry and ion-ion interactions near interfaces, may not be captured using a fixed cutoff distance method. Further developments in the treatments of electrostatics for these kinds of systems should be made in order to render the method suitable for charge-dependent interfaces.

\subsection{Force and energy cap}

So far most of the H-AdResS applications focused on bulk liquids \cite{ Potestio2013, Heidari2016}. In the CG region, every liquid molecule was mapped to a single CG bead that only interacts with other beads through non-bonded interactions. Thus, bonded interactions between CG beads such as bonds, angles, proper and improper dihedrals, were not needed. Since the topology in the CG region is unaltered and atomistic degrees of freedom are preserved everywhere, the total number of particles in the simulation box remains constant. However, in cases where we have more than one CG bead per molecule in the CG region, the CG description will result in a different topology compared to that of the AA resolution. In the current AdResS and H-AdResS implementations \cite{agromacs, aespres, Heidari2016}, bonded interactions are computed considering the underlying atomistic DOFs in the whole simulation box.

\begin{figure}[t]
    \centering
    \includegraphics[width=0.5\textwidth]{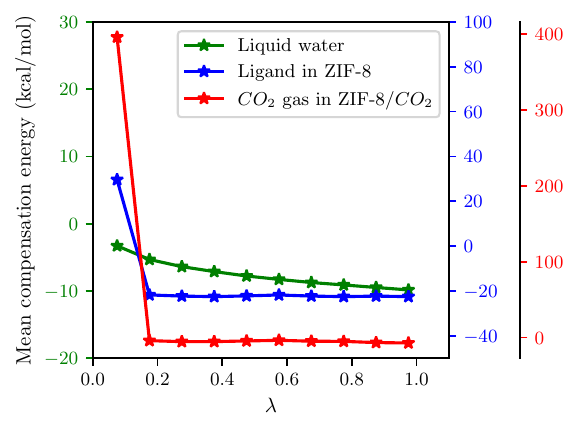}
    \caption{Average compensation energy applied in the H-AdResS simulation of liquid water (green curve), bulk ZIF-8 (blue curve) and ZIF-8/\ce{CO2} (red curve) without the energy and force capping mechanism plotted against the resolution of the particle in the hybrid region.}
    \label{fig:before-cap}
\end{figure}

During the simulation, there may be configurations in which molecules lie in-between the CG and HY regions. Since their non-bonded interactions are governed by CG force fields that are applied to their center of mass and redistributed by mass fraction, the underlying atoms may reach distances that would not be possible in an atomistic simulation (See Figures \ref{fig:scheme1} and \ref{fig:scheme2}). Due to thermal fluctuations, these molecules in the boundaries of the CG region can oscillate between the CG and HY region. This results in an intermittent activation and deactivation of atomistic non-bonded DOFs, which contributes to non-physical energies and forces to the Hamiltonian. Since the compensation forces are calculated as a difference between the potential energies of a molecule in the AA and CG regions (see eq. \ref{eq:force}), these non-physical energies and forces translate into the pressure compensation scheme. In Figure \ref{fig:before-cap}, we compare the mean compensation energy applied on water molecules (green curve), ligands of ZIF-8 (blue curve), and \ce{CO2} gas in ZIF-8/\ce{CO2} (red curve) respectively. This quantity changes smoothly as a function of $\lambda$ for water. But for ZIF-8 a sharp peak is observed for the first bin at the boundary of HY and CG regions ($\lambda \xrightarrow{} 0$). This peak grows taller for the complex heterogeneous case of \ce{CO2} in ZIF-8. 

\begin{figure}[h]
    \centering
    \includegraphics{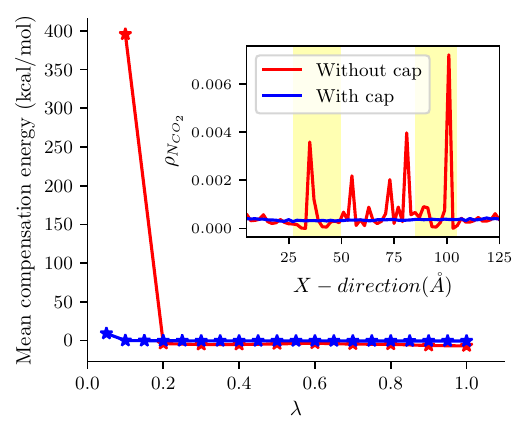}
    \caption{Average compensation energy applied to the \ce{CO2} molecules in the hybrid region plotted against the resolution of the particle ($\lambda$). The subplot represents the number density profile of \ce{CO2} gas in the H-AdResS simulation with pressure and density compensation routines active. The red and blue curves were obtained from H-AdResS simulations without and with force and energy capping mechanisms respectively. The yellow stripes in the subplot denote the position of the HY regions in the simulation box.}
    \label{fig:w_wo_c}
\end{figure}

Since ZIF-8 is a bulk solid, the impact of these non-physical forces and energies acting on it is mitigated by vibrations through the network and does not result in structural defects. However, when such high force is applied on the gas molecules in the HY region, they tend to move selectively in the direction in which the force is applied and accumulate there. Though pressure and density compensation routines are activated, these non-physical forces cause an imbalance in the density of the gas. In some cases, this even leads to crashing simulations.
Intuitively, we could think that this problem can be circumvented just by capping these spurious forces acting over the particles. However, eq. \ref{eq:force} shows that the force introduced for compensating the difference in pressure between AA and CG regions is directly proportional to the energies. Therefore, capping energies is also needed. For these reasons, our version of H-AdResS presented in Ref. \citenum{Sudhakar2026} implements a hard cap on both the forces and energies in the hybrid region.

In our case, the energies and forces are capped at 1000.0 kcal/mol and 1000.0 kcal/(mol \AA) respectively. Figure \ref{fig:w_wo_c} compares the compensation forces applied on the gas molecules and the density profile of \ce{CO2} gas in H-AdResS simulations with pressure and density compensation, in the presence and absence of force and energy capping respectively. It is clear that in the absence of a capping mechanism, the gas molecules experience a non-physical drift despite the presence of compensation routines. Thus, capping forces and energies in the hybrid region is essential for simulating systems that contain more than one CG bead mapping such as ZIF-8 and other extended solids, in the CG region  of H-AdResS simulations. In our case, on average $\sim$35 capping events occur for every nanosecond of a system containing $\sim$9000 atoms, which roughly corresponds to $0.00000001\%$ of the total number of pairwise interactions calculated. Therefore, the capping scheme is only applied very rarely and does not perturb the structure and dynamics of the system.

\subsection{Inclusion of the CG bonded terms in the Hamiltonian}

Finally, the most important challenge of the H-AdResS scheme is the treatment of bonded interactions with the underlying atomistic degrees of freedom throughout the whole simulation box, including the CG region. Even though this facilitates the seamless change in resolution by circumventing the problem of changing the total number of particles in the simulation box on-the-fly, this methodology poses challenges. CG force fields are typically parameterized to reproduce either structural properties or the free energy surface of a benchmark AA model.\cite{Rzepiela2011, FalcnGonzlez2020, Agarwal2025, Carbone2008} They have a simpler topology and inherit a smoother free energy surface compared to AA force fields. Even though CG force fields can be parameterized well enough to reproduce key properties of the AA force fields, this is only guaranteed when they are used in their full form, as developed. Since the AdResS and H-AdResS schemes so far demand using the underlying AA DOFs instead of the CG bonded interactions in the whole simulation box, we are not employing the accurate equation of state in the CG region according to the CG force field. 

Molecules in the CG region rely actively on the AA bond, angle, torsion and improper dihedral constraints but do not carry the respective non-bonded AA terms. Therefore, the underlying inactive degrees of freedom evolve according to force fields that are inconsistent with their CG non-bonded interactions. This means that molecules in the CG region are treated by their center of mass for their non-bonded interactions but constrained with AA bonded DOFs (see Figure \ref{fig:mapping}), resulting in non-physical mixed states such as those illustrated in Figure \ref{fig:scheme1} and \ref{fig:scheme2} and quantified in Figure \ref{fig:pair-dist}. Though the center of mass distance between the ligands, and between ligand and gas are within what is expected for the non-bonded CG force fields, the distances between underlying atoms can become non-physical. This makes any structural parameter measured in the CG region unreliable, even though the original CG force fields that contain bonded contributions adapted to the CG topology are capable of reproducing those parameters.\cite{Alvares2025} Since the bonded terms appear in the Hamiltonian explicitly, removing them from the CG region would make the Hamiltonian invalid.

We made an attempt to include the CG bonded terms in the CG region. To this end, we developed new bond and angle styles to treat the bonded interactions in the AA and CG regions respectively. For the AA region, bond styles such as \textit{harmonic/hars} (\texttt{Class BondHarmonicHARS}), angle styles such as \textit{harmonic/hars} (\texttt{Class AngleHarmonicHARS}) and \textit{ charmm/hars} (\texttt{Class AngleCharmmHARS}), dihedral style \textit{harmonic/hars} (\texttt{Class DihedralHarmonicHARS}) and improper style \textit{cvff/hars} (\texttt{Class ImproperCvffHARS}) were developed. For the CG region, bond style \textit{table/hars} (\texttt{Class BondTableHARS}) and angle style \textit{table/hars} (\texttt{Class AngleTableHARS}) were developed respectively. All these new styles are linked with the parent H-AdResS \texttt{class FixLambdaHCalc}. This gives access to the $\lambda$ values of the atoms to these new pair styles, thus enabling the handling of bonded interactions following the region-specific force field definition.

With the newly developed styles, AA bonded terms were removed and compensated with the CG bonded terms in the CG region. The total force due to the bonded and non-bonded interactions experienced by molecules in the CG region was calculated using their center of mass distances. Similar to non-bonded forces, the bonded forces were also distributed to underlying atoms based on their mass fraction. The updated Hamiltonian takes the form:
\small
\begin{equation} 
H  = \kappa + \sum_a \{\lambda_a (U_{bonded}^{AA} + U^{AA}_a) + (1-\lambda_a)(U_{bonded}^{CG} + U^{CG}_a) \}
\label{eq:up-hamiltonian}
\end{equation} 
\normalsize

\begin{figure}
    \centering
    \includegraphics[width=0.5\textwidth]{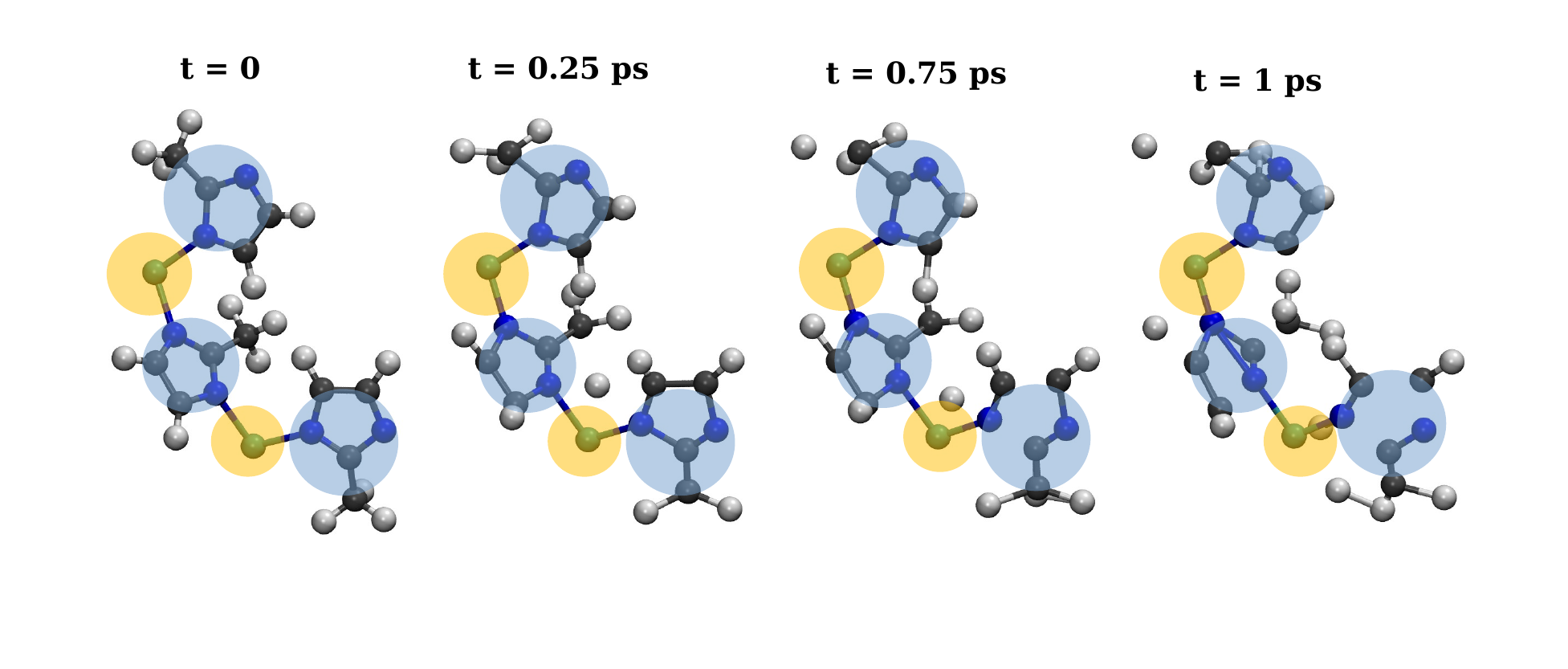}
    \caption{Schematic representation of ZIF-8 simulated in the H-AdResS scheme with an updated Hamiltonian given by the eq. (\ref{eq:up-hamiltonian}). Carbon atoms are denoted in grey, Nitrogen atoms in blue and Hydrogen atoms in white color. The blue and orange sphere represent the CG mapping of the ZIF-8 ligand and metal species.}
    \label{fig:hexane}
\end{figure}

Figure \ref{fig:hexane} represents a snapshot from a H-AdResS simulation of ZIF-8 using the updated Hamiltonian (given in eq. \ref{eq:up-hamiltonian}). The t=0 panel shows part of the topology of the ZIF-8 molecule at the beginning of the simulation. The colored spheres denote the CG mapping. With time, strong perturbations in the underlying atomistic topology can be observed, though the CG mapping looks intact. The flying hydrogen atoms are loosely held due to the lack of atomistic forces and lead to crashing simulations. Though every atom experiences a net force, these forces are obtained from considering the CG level mapping but not accounting for their underlying atomic neighbors. Thus the net force experienced by the underlying atoms does not correspond to the potential of mean force due to their atomic neighboring environment. This leads to the destruction of the underlying atomistic topology. 

This makes it technically impossible to remove the AA bonded terms while the underlying atomistic degrees of freedom exist. The only way of circumventing this problem would be to remove the underlying atoms in the CG region and adapt the number of particles in the simulation box on-the-fly accordingly. Still, treating molecules in the hybrid region would be challenging, because of the fractional degrees of freedom possessed by the molecule, and the fact that the entropy changes during the dynamic change in number of particles cannot be compensated with a thermostat. 

Even in the existing formulation, the term $\Delta H(\lambda(X_{a}))$ is parameterized in such a way that it compensates the free energy difference between the AA and CG regions. However for a canonical ensemble, the free energy difference contains not only energy but also entropy contributions : $\Delta A = \Delta U - T\Delta S$. Though the number of atoms are constant, the time average of $\langle U^{AA} - U^{CG}\rangle$ applied on the particles in the HY region as a function of $d\lambda$, does not capture the changes in the entropy due to the activation and deactivation of the degrees of freedom. This formulation may ensure uniform density and pressure on average. However, the unaccounted change in entropy during the resolution change will perturb the canonical distribution of molecules in the AA region during re-insertion.

Furthermore, every time that a molecule enters the AA region, all of its DOFs are fully activated. The atomistic topology with which the molecule enters the AA region carries the memory of the dynamics that it experienced in the CG region. Thus, the orientation in which the molecule enters the AA region may not be consistent with the canonical distribution reproduced by the atomistic force fields in the AA region. For systems with very high relaxation times, re-introducing the degrees of freedom in such a non-equilibrated fashion can introduce artifacts that are tenacious and cannot be averaged out statistically over the timescale of the simulation. This impacts molecules dynamics in short timescales.

Deactivating the underlying atomistic degrees of freedom in the CG region, together with the non robust handling of entropic contributions in compensations lead to sampling a different statistical ensemble in the AA region of a H-AdResS simulation than in a fully atomistic simulation. The discrepancies are small, so that the equilibrium structure and dynamic properties that are measured as an average over many molecules are still consistent. Nevertheless systems with very high relaxation times or strong correlations between DOFs require an improved handling of phase space coupling, especially in the resolution boundaries within the current H-AdResS framework.

\section{Conclusions}

In this work, we first provide a comprehensive analysis on thermostatting and electrostatics handling methods applied to H-AdResS simulations. We simulated two systems representing pure liquids and solid/gas interfaces: liquid water and ZIF-8/\ce{CO2} within the H-AdResS scheme using Nosé-Hoover and Langevin thermostats.

Atomic temperature profiles revealed that the use of the Nosé-Hoover thermostat results in a thermal gradient among different atom-types. Due to its global behavior, it fails to restore the energy equipartition perturbed by the local fluctuations that arise from changes in the resolution. This leads to an inaccurate structure in the AA region and artifacts such as the flying ice cube effect. On the other hand, using the Langevin thermostat is effective in maintaining an uniform temperature distribution among different atom-types, making it the right choice for H-AdResS simulations. We recommend using two Langevin thermostats, one for the AA region and another for the hybrid and CG regions, to maintain uniform temperature in the different box regions via the choice of appropriate friction terms for each region.

The DSF method can yield accurate stucture and dynamics in the AA region of H-AdResS for bulk systems. To understand the possibilities of studying charged interfaces using this method within the H-AdResS scheme, we studied its suitability to reproduce properties that are sensitive to electrostatic interactions. We compared the dipolarorientation profile of a water-vapour interface using either the PPPM or the DSF method. The DSF method failed to reproduce the structure and polarization of water molecules at the interface. This suggests that using the DSF method within the H-AdResS scheme would not be suitable for systems involving charged interfaces. 

H-AdResS simulations involving molecules mapped to more than one CG bead introduced non-physical forces and energies at the boundaries between the hybrid and CG regions. This can hinder the dynamics and even lead to the crashing of simulations. To handle this, energy and force capping mechanisms are needed. This is illustrated by plotting the compensation energies and density profile of the diffusing \ce{CO2} gas in ZIF-8 with and without the capping mechanism.

Finally, the fundamental H-AdResS formulation and the challenges involved such as underlying atomistic degrees of freedom and lack of an entropic contribution in compensations were discussed. Implementing the CG bonded terms in the CG region fails to preserve the underlying atomistic topology. Dynamic number of particles could solve the problem, but the fractional degrees of freedom in the hybrid region and entropic changes still pose challenges. Within the current scheme, only the AA region can yield valid and physically accurate structure and dynamics. The CG region does not represent the true CG behaviour because of the mixed force field representation due to the underlying atomistic topology. Thus, quantities measured in the CG region are not valid and may not reproduce true CG structure and dynamic behaviours. As long as we are only interested in the AA region, the use of the HAddress scheme is appropriate. This aligns exactly with the scope for which the HAddress scheme is intended to be used.

H-AdResS is a powerful method that offers an adaptive framework with a Hamiltonian formulation, for molecules to change their resolution on-the-fly during the simulation. The scheme is useful in studying various bulk and heterogeneous systems with a computational speedup compared to fully atomistic simulations. Nevertheless, using the method requires care and taking into account its limitations. We hope that this work will help people determining whether this method is useful for their target application and how to make it work successfully.

\section*{Acknowledgements}

The authors thank the Ecole Doctorale Chimie Physique et Chimie Analytique de Paris Centre for funding this work. R.S. thanks the European Research Council for an ERC StG (MAGNIFY project, number 101042514).  This work was granted access to the HPC resources of the SACADO MeSU platform at Sorbonne Université, where the simulations were performed. 


\section*{Conflict of Interest}
The authors have no conflicts to disclose.



\section*{Supporting information}

Additional details on symbols and mapping schemes of ZIF-8 and \ce{CO2}, temperature profiles of different regions and species computed using one and two Langevin thermostat and snapshots of underlying atomistic DOFs in the boundary of HY and CG regions are provided in the supporting information.

\section*{References}
\bibliography{references}

\clearpage

\onecolumngrid

\setcounter{equation}{0}
\setcounter{figure}{0}
\setcounter{table}{0}
\renewcommand{\theequation}{\roman{equation}}
\renewcommand{\thefigure}{S\arabic{figure}}
\renewcommand{\thetable}{S\arabic{table}}
\renewcommand{\thepage}{S\arabic{page}}

\section*{Supporting information}

\begin{figure}[h]
    \centering
    \includegraphics[width=0.5\textwidth]{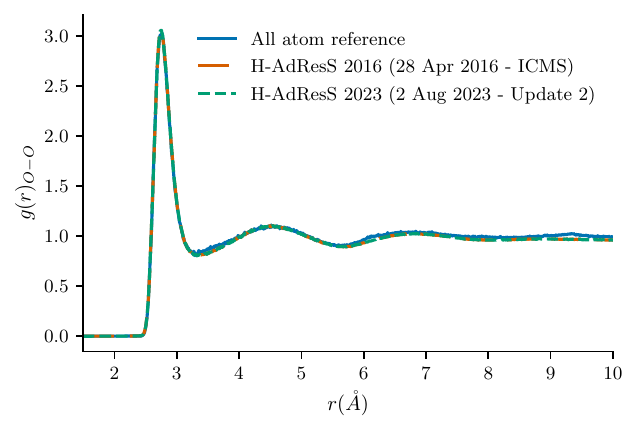}
    \caption{Radial distribution function, g(r), of the O--O pair in bulk liquid water computed from a full AA simulation [performed with Nosé-Hoover thermostat] (blue solid line) used as a reference against the g(r) computed in the AA region of a H-AdResS simulations [performed with langevin thermostat] performed using the 2016 implementation (orange long dashes) and our 2023 implementation (green dashes). H-AdResS simulations are performed with both the compensations active. }
    \label{fig:rdf-water}
\end{figure}

\begin{figure*}[h]
    \centering
    \includegraphics[width=0.45\textwidth]{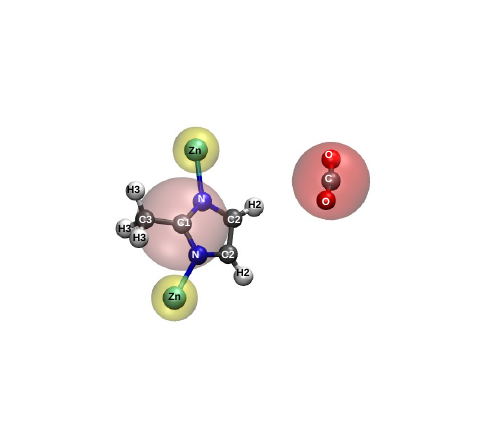}
    \caption{Schematic representation of atomistic and coarse-grained mapping of ZIF-8 moieties and \ce{CO2} respectively. In the atomistic scheme, Carbon atoms are colored in grey, Hydrogen atoms in white, Nitrogen atoms in blue, Zinc atoms in cyan and Oxygen atoms in red. In the coarse-grained scheme, zinc is mapped into a yellow bead, the whole 2-methyl imidazolate ligand is mapped into a pink bead and \ce{CO2} into a red bead. Transparent spheres denoting the CG representation of the molecule are used to calculate the non-bonded interactions, whereas the underlying atomistic degrees of freedom are held together by the bonded terms from the atomistic force field. } 
    \label{fig:mapping}
\end{figure*}

\begin{figure*}
    \centering
    \includegraphics[width=0.8\textwidth]{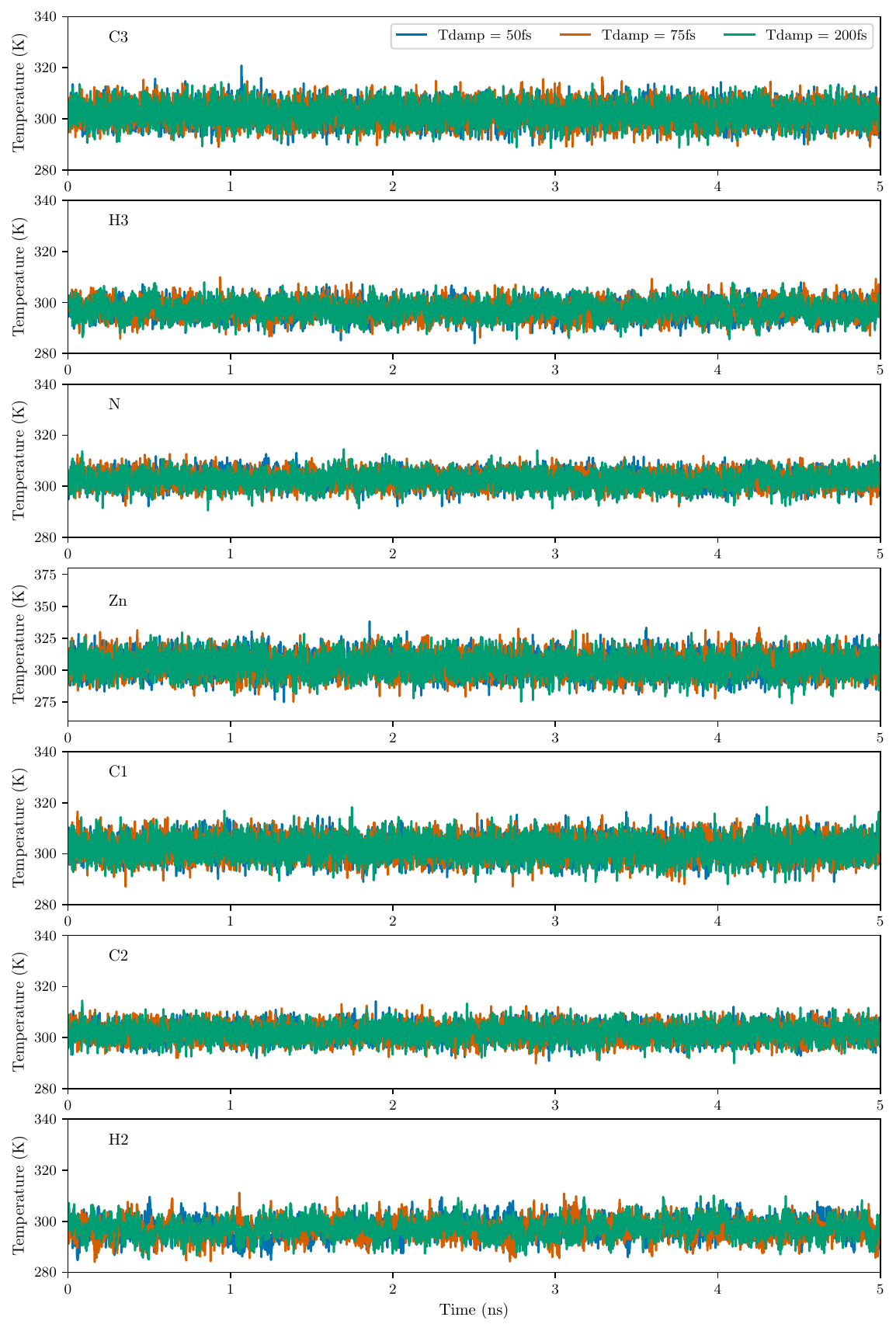}
    \caption{Temperature of different atom-types of ZIF-8 computed with Nosé-Hoover thermostat using three different damping constant (Tdamp) namely 50fs (blue curve), 75fs (orange curve) and 200fs (green curve) respectively. The simulations are performed using LAMMPS 2023 implementation reported in the ref. \citenum{Sudhakar2026}. }
    \label{fig:damping}
\end{figure*}

\begin{figure*}
    \centering
    \includegraphics[width=0.8\textwidth]{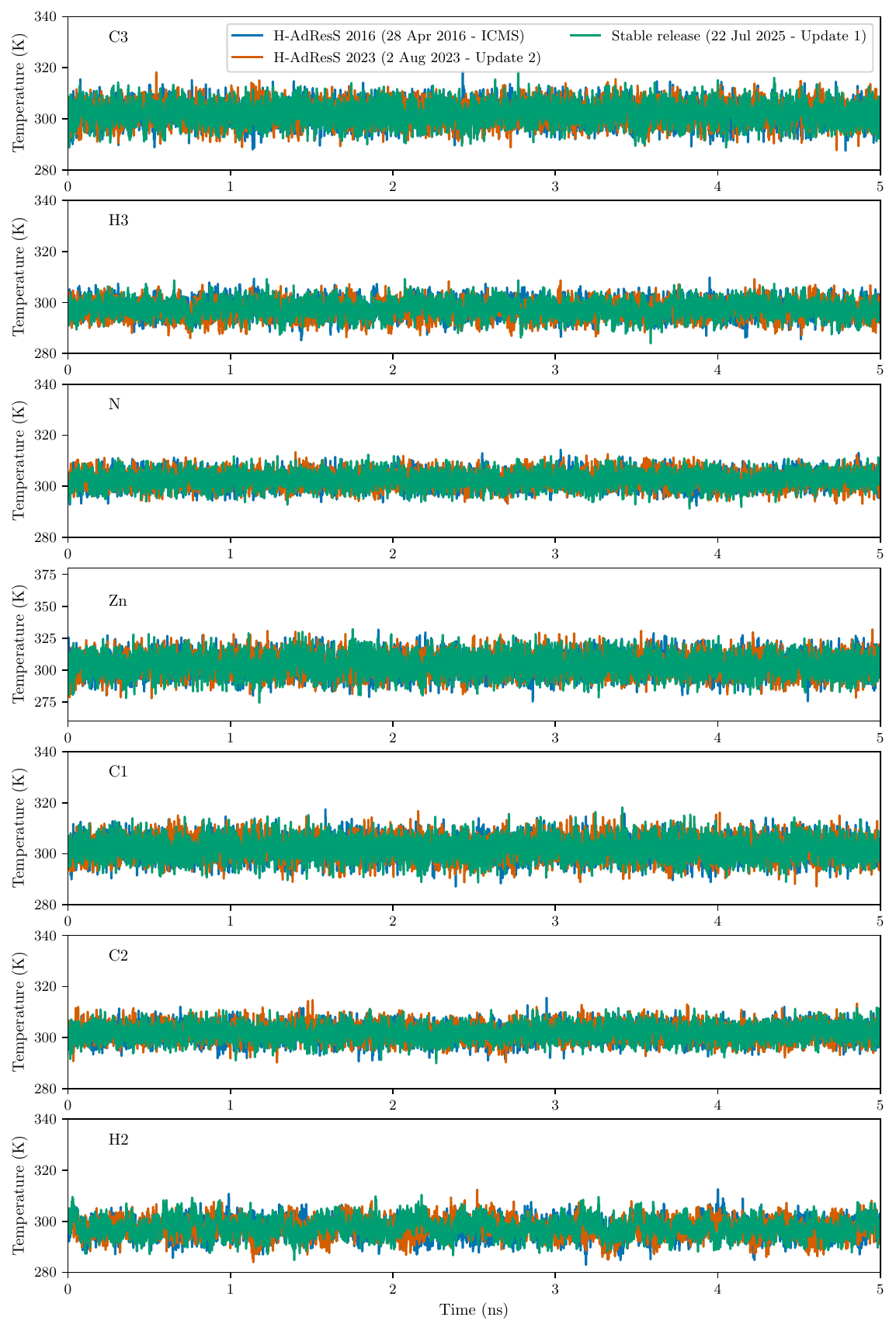}
    \caption{Temperature of different atom-types of ZIF-8 computed with Nosé-Hoover thermostat using three different LAMMPS implementation.}
    \label{fig:version-thermo}
\end{figure*}

\begin{figure*}
    \centering
    \includegraphics[width=0.8\textwidth]{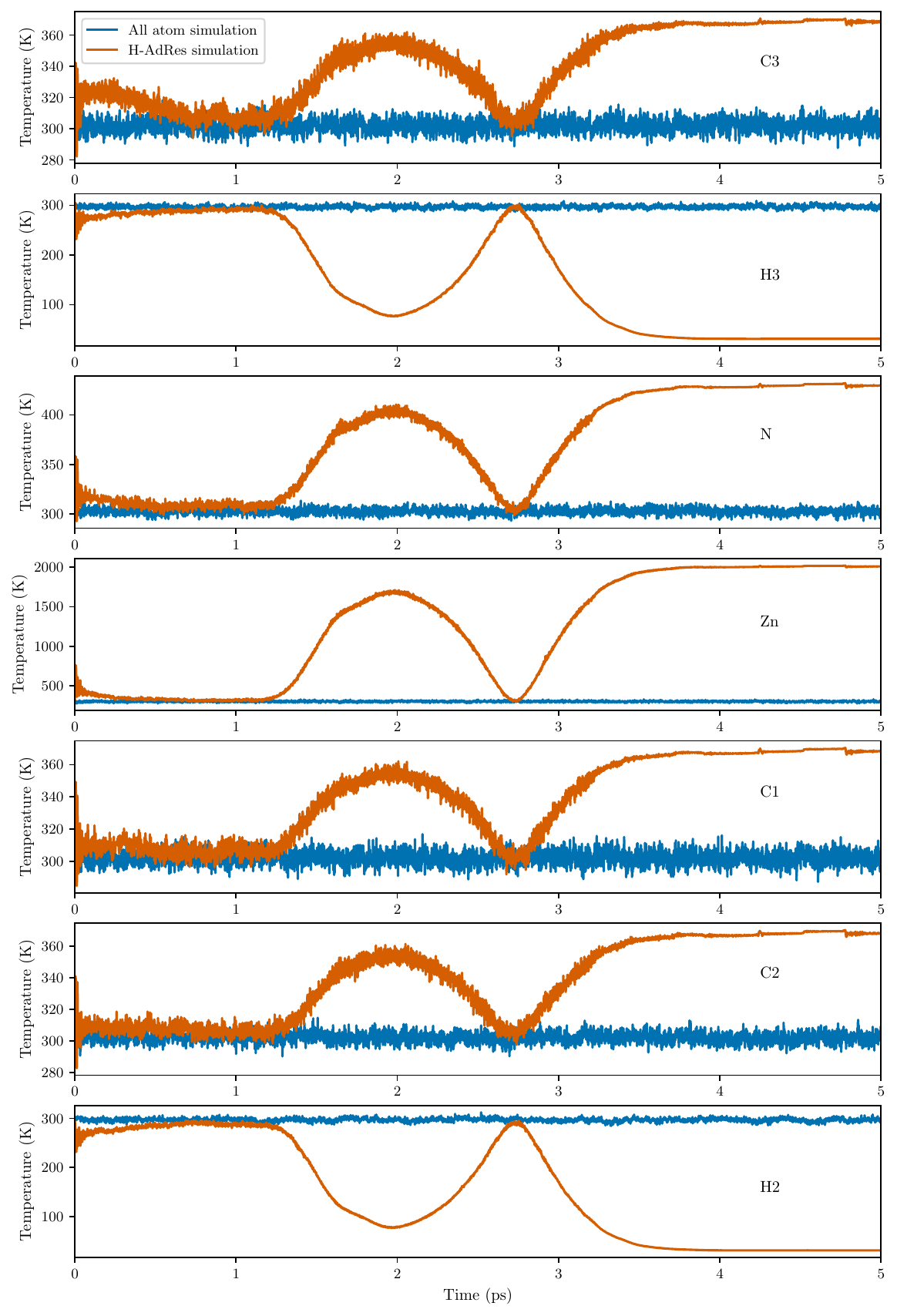}
    \caption{Temperature of different atom-types of ZIF-8 computed from an all atom simulation (blue curve) and H-AdResS simulation (orange curve) using Nosé-Hoover thermostat. The simulations are performed using LAMMPS 2023 implementation reported in the ref. \citenum{Sudhakar2026}.}
    \label{fig:aa-vs-hars}
\end{figure*}

\begin{figure*}
    \centering
    \includegraphics[width=0.9\textwidth]{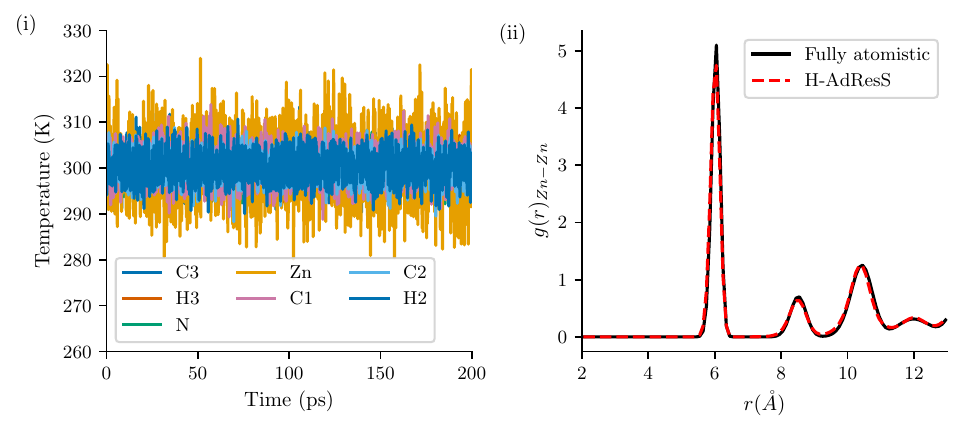}
    \caption{ (i) Temperature of different atom-types of ZIF-8 and (ii) radial distribution function g(r) of Zn--Zn pairs, computed in the AA region of the H-AdResS simulation with both compensations active using two Langevin thermostats.}
    \label{fig:two-ln}
\end{figure*}

\begin{figure*}
    \centering
    \includegraphics{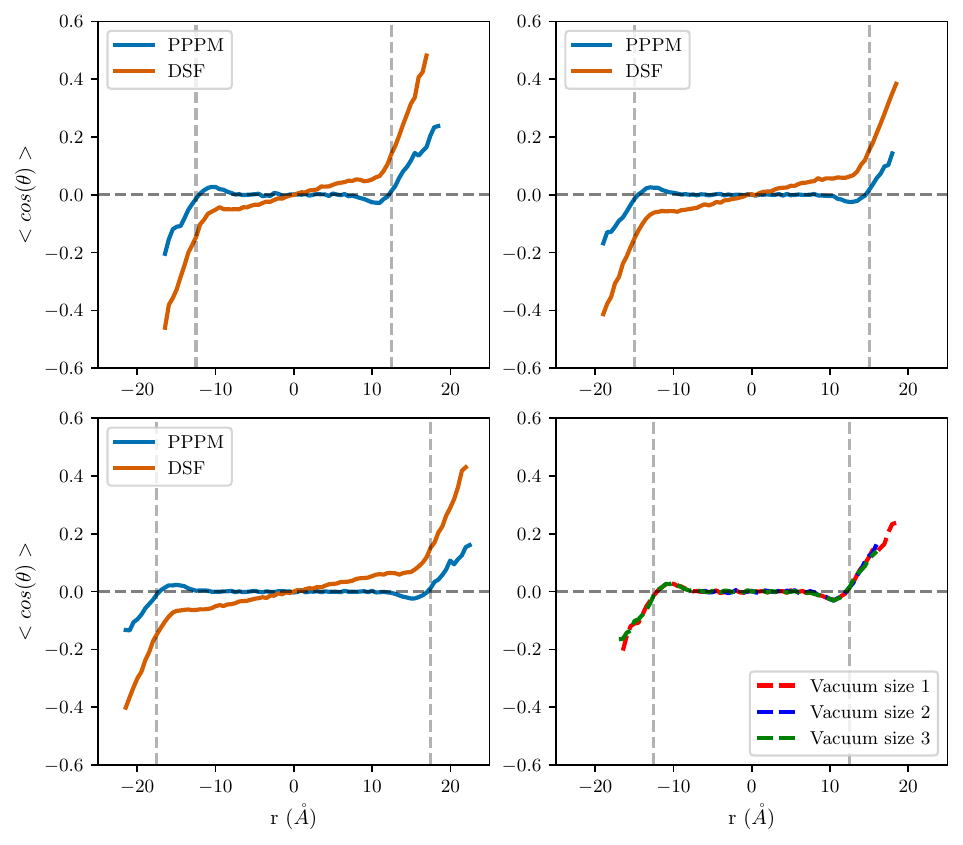}
    \caption{Average dipolar orientation profile computed for liquid water slab forming an interface with vacuum. 3 different thickness of water slabs were simulated using PPPM (blue curve) and DSF (orange curve) methods respectively. The slab thickness were 20 \AA (top left), 25 \AA (top right) and 30 \AA (bottom left), while other two dimensions of the water slab are kept fixed (20 \AA x 20 \AA). Bottom right panel compares the dipolar orientation profile computed for the same water slab but with different vacuum region sizes. The vertical dotted lines denote the position of the liquid water slab.}
    \label{fig:dipole-profiles}
\end{figure*}

\begin{figure*}
    \centering
    \includegraphics[width=0.9\textwidth]{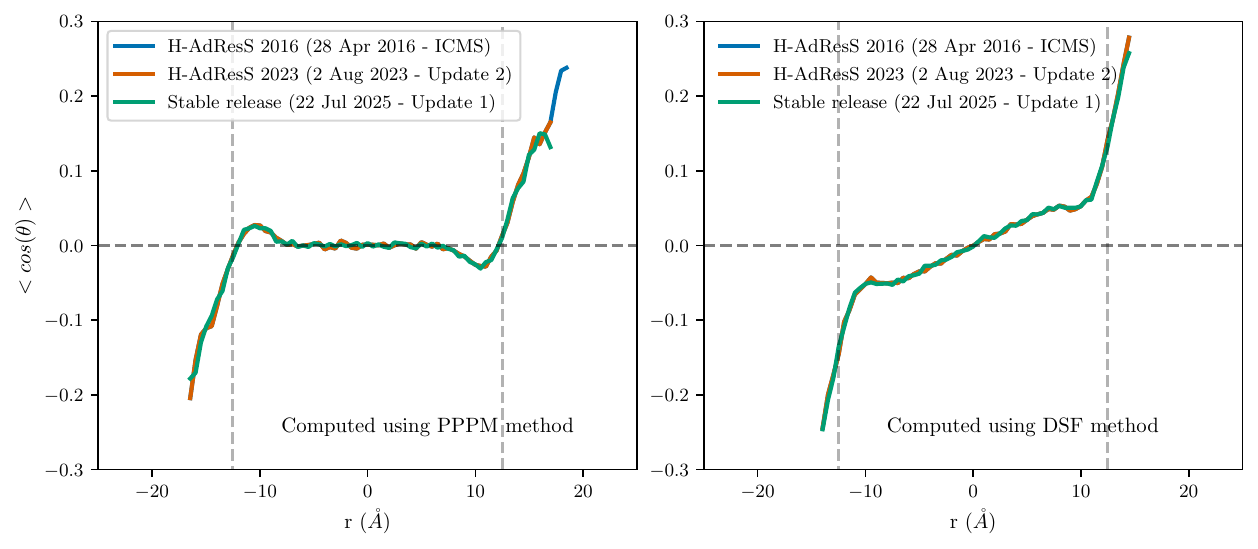}
    \caption{Average dipolar orientation profile computed for liquid water slab forming an interface with vacuum computed with 3 different versions of LAMMPS, using PPPM (left panel) and DSF (right panel) methods respectively. The vertical dotted lines denote the position of the liquid water slab.}
    \label{fig:my_label}
\end{figure*}

\begin{figure*}
    \centering
    \includegraphics[width=0.45\textwidth]{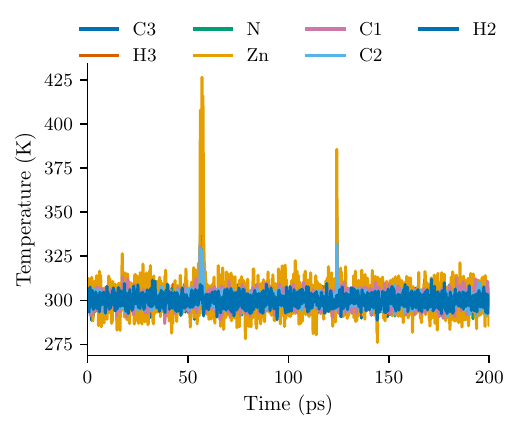}
    \caption{ Temperature of different atom-types of ZIF-8 computed in the AA region of the H-AdResS simulation with both compensations active using one Langevin thermostat.}
    \label{fig:one-ln}
\end{figure*}

\begin{figure*}
    \centering
    \includegraphics[width=0.45\textwidth]{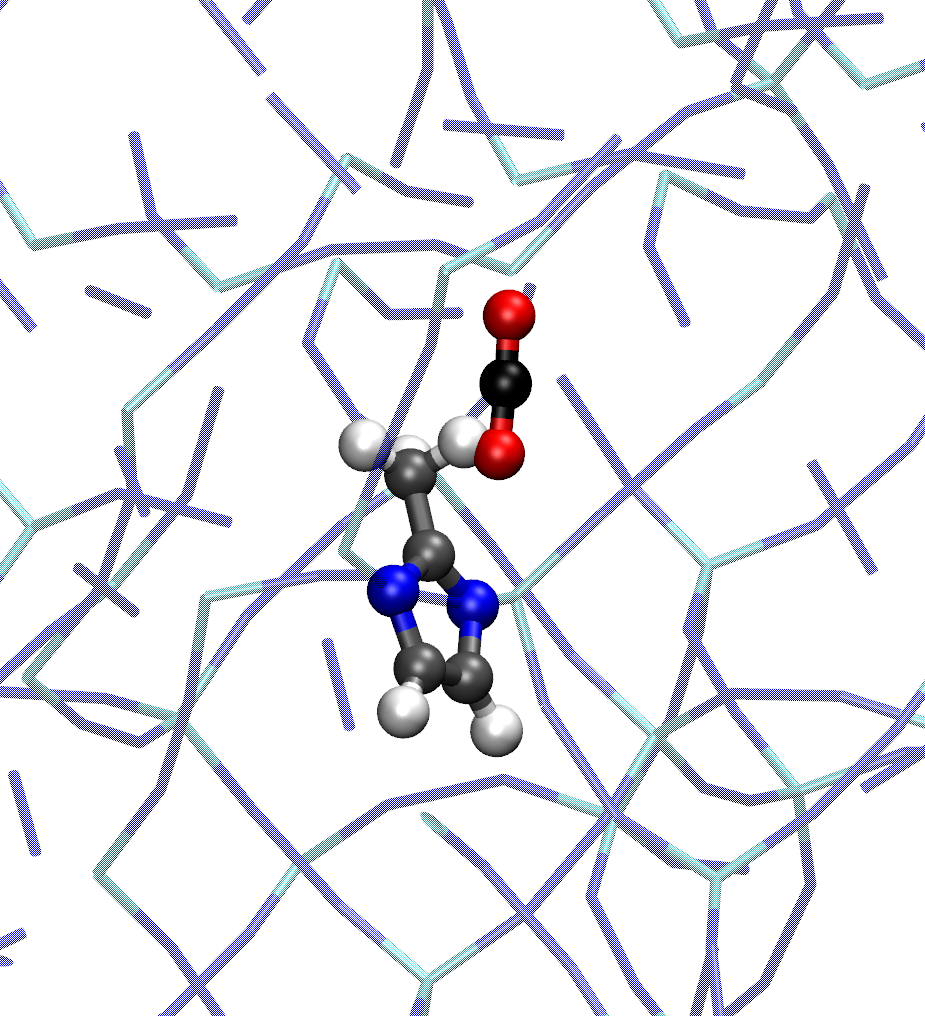}
    \caption{Visual representation of the underlying atomistic degrees of freedom of ZIF-8 and \ce{CO2} in the boundary of HY and CG regions attaining a configuration that is not possible in the fully atomistic simulation.}
    \label{fig:scheme1}
\end{figure*}

\begin{figure*}
    \centering
    \includegraphics[width=0.45\textwidth]{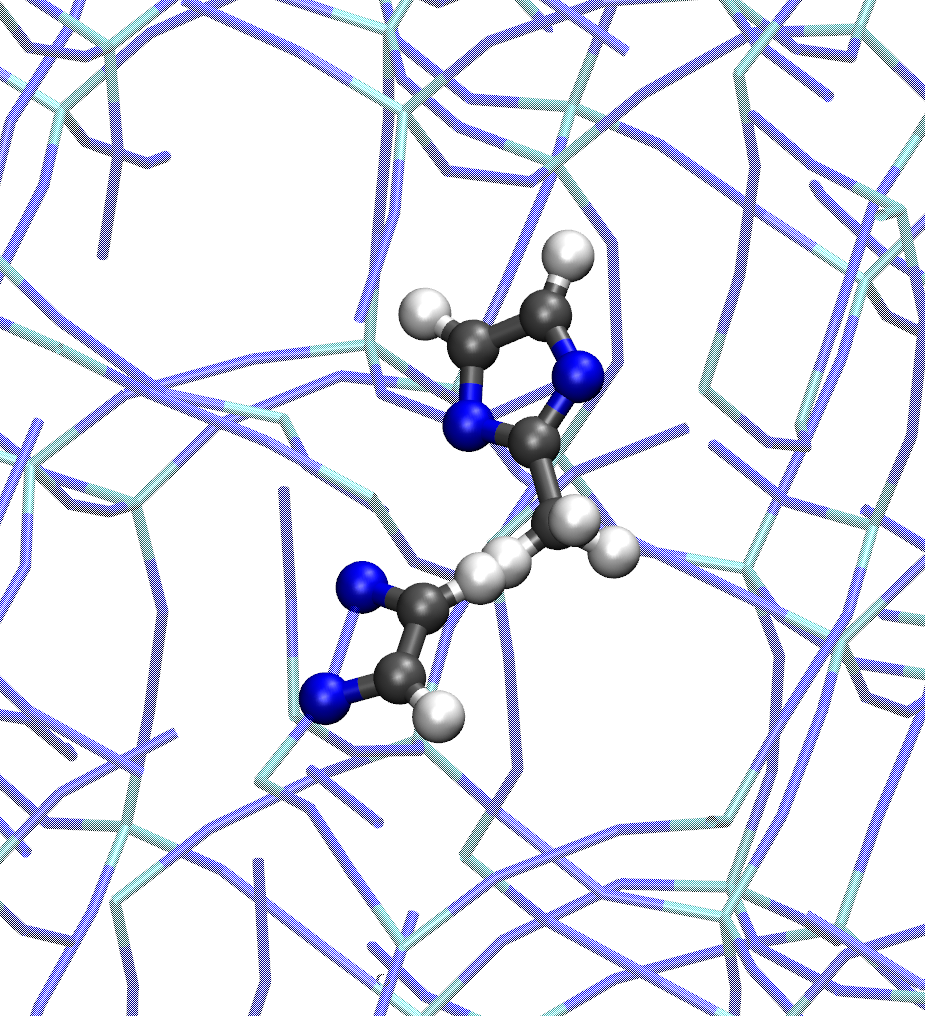}
    \caption{Visual representation of the underlying atomistic degrees of freedom of ZIF-8 in the boundary of HY and CG regions. Two methylimidazolate ligands attaining a configuration that is not possible in the fully atomistic simulations.}
    \label{fig:scheme2}
\end{figure*}

\begin{figure*}
    \centering
    \includegraphics{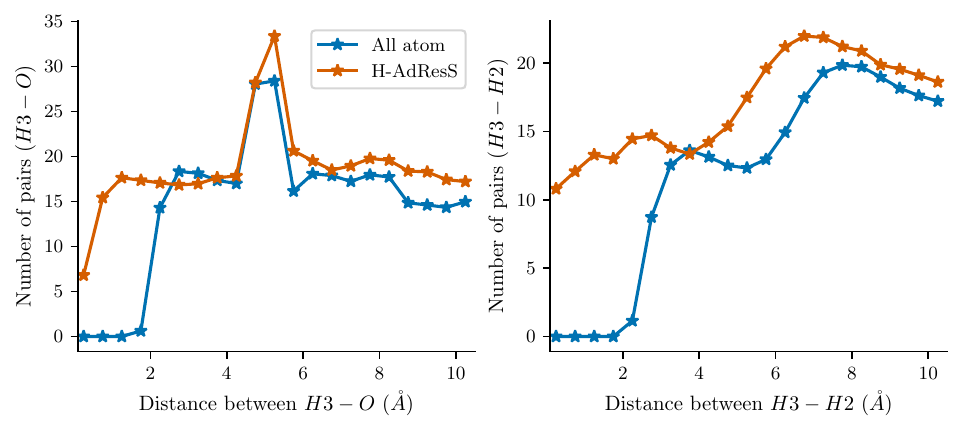}
    \caption{Quantitative analysis of number of H3 (ZIF-8) - O (\ce{CO2}) and H3 (ZIF-8) - H2 (ZIF-8) atom pairs and the distance between them, in the boundary of hybrid and coarse-grained region.}
    \label{fig:pair-dist}
\end{figure*}

\end{document}